\documentclass[twocolumn,showpacs,preprintnumbers,amsmath,amssymb]{revtex4}

\usepackage{graphicx}
\usepackage{dcolumn}
\usepackage{bm}

\begin{document}

\title{Time-Dependent Quantum Weak Values: \\ Decay Law for Post-Selected States}
\author{P.C.W. Davies}
\affiliation{The Beyond Center for Fundamental Concepts in Science, Arizona State University, Tempe, AZ 85287-1504}
\date{\today}

\begin{abstract} Weak measurements offer new insights into the behavior of quantum systems. Combined with post-selection, quantum mechanics predicts a range of new experimentally testable phenomena. In this paper I consider weak measurements performed on time-dependent pre- and post-selected ensembles, with emphasis on the decay of excited states. The results show that the standard exponential decay law is a limiting case of a more general law that depends on both the time of post-selection and the choice of final state. The generalized law is illustrated for two interesting choices of post-selection. \end{abstract}

\pacs{03.65.-w, 03.65.Ca, 03.65.Ta}

\maketitle


\section{BACKGROUND}

Standard quantum mechanics famously describes two varieties of time evolution: the normal unitary evolution of the wave function, and the abrupt non-unitary ÒcollapseÓ occasioned by measurement, when the system is projected onto an eigenstate of the relevant observable. The latter process involves an irreversible disturbance of the quantum system which alters its further evolution. Von Neumann's analysis of the measurement process, in which the quantum system of interest is strongly coupled to an external measuring device, explicates this non-unitary projection and unavoidable disturbance \cite{bib1}. A significant modification to this simple picture emerges, however, if one considers a quantum measurement carried out on an ensemble of systems.

Consider a collection of $N$ non-interacting systems prepared initially in the product state

\begin{equation}\label{Eq_1_}
|\Psi^{(N)}\rangle = |\psi\rangle_1 |\psi\rangle_2 |\psi\rangle_3 \ldots |\psi\rangle_N
\end{equation}
and a set of identical observables $\{A_j\}$, where  $A_j$  acts on the $j$th member of the ensemble, with corresponding Hermitian operators $\hat{A}_j$. The ensemble average operator may be defined as 

\begin{equation}\label{Eq_2_}
\hat{A}^{(N)} = (1/N) \sum_{j=1}^N \hat{A}_j
\end{equation}
from which it is obvious that the expectation value of the average $\langle\hat{A}^{(N)}\rangle$ is the same as the average of the expectation values  $\langle \hat{A}_j \rangle$. Let us now restrict to the special case that the $N$ systems are identical, and prepared in the same initial state $|\psi\rangle$, which may be achieved, for example, by making a strong, projective measurement of the same observable on every member of a large ensemble, and retaining in a sub-ensemble only those members that satisfy the required initial condition. It may then readily be proved \cite{bib3b} that in the limit $N \to \infty$

\begin{equation}\label{Eq_3_}
\hat{A}^{(N)} |\Psi^{(N)}\rangle \to \langle\hat{A}\rangle|\Psi^{(N)}\rangle
\end{equation}
where $\langle\hat{A}\rangle$ is the expectation value for a \emph{single} member of the ensemble (the running index has been suppressed because all members are identical).  Furthermore, a measurement of $A^{(N)}$ will not disturb the system, or any individual member thereof, in the large $N$ limit.  The foregoing properties are supported by detailed calculations of the actual measurement process, for example, based on von Neumann's model of a measuring device coupled to the entire ensemble \cite{bib2}.  Thus the measurement yields non-trivial new information about the system, namely $\langle\hat{A}\rangle$, without (in the large $N$ limit) disturbing the measured system at all.  This result, which is contrary to the usual wisdom, may be understood intuitively by noting that $\langle\hat{A}\rangle$  enters into Eq.\eqref{Eq_3_} as an \emph{eigenvalue} of $\hat{A}^{(N)}$ (though it is not in general an eigenvalue of $\hat{A}$).  It is well known that repeated measurement of an eigenstate does not disturb the measured system.

The key to understanding the foregoing result is readily apparent from inspecting Eq.\eqref{Eq_2_}. If the measuring device couples to the whole ensemble with fixed strength, then the coupling to any \emph{individual} member of the ensemble is reduced by the pre-factor of the right hand side, $1/N$. When $N$ is very large, the coupling to individual systems is very weak, and in the limit $N\to\infty$ the coupling approaches zero. Weak measurements imply small disturbance on the measured system, and in the limit being discussed, the disturbance will be arbitrarily small. It can be explicitly verified \cite{bib2} that the probability of a measurement disturbing \emph{any} member of the ensemble approaches zero like $1/N$.  Nevertheless, information about the \emph{average} is acquired.

One may now ask what would be the outcome of a \emph{strong} measurement performed on a given member $i$ of the ensemble \emph{after} the weak measurement has been performed to disclose some information about $i$.  Although each member of the ensemble is prepared in an identical state, subsequent strong measurements on $i$ will not, of course, generally yield the same eigenvalue; rather, eigenvalues will be distributed with relative probabilities according to the Born rule.  The initial ensemble may therefore be split into sub-ensembles according to the outcomes of the final strong measurement, and we are free to focus on a specific sub-ensemble.  In other words, in addition to pre-selecting an ensemble of $N$ identically-prepared systems, we may also \emph{post-select} a sub-ensemble of $\leq N$ systems satisfying a final condition too. In this paper, I shall consider the statistics of weak measurements carried out on specific pre- and post-selected sub-ensembles at intermediate times, i.e. \emph{after} the (strong) pre-selection process and \emph{before} the (strong) post-selection process.

If the initial and final states of the whole ensemble are denoted $|\Psi_i\rangle$ and $|\Psi_f\rangle$ respectively, and the corresponding single-member states by $|\psi_i\rangle_j$ and  $|\psi_f\rangle_j$, then the expectation value of $\hat{A}$ for the $j$th member of the ensemble in the initial state may be decomposed as follows:

\begin{equation}\label{Eq_4_}
\langle \Psi_i |\hat{A}_j| \Psi_i \rangle = \sum_k | \langle \Psi_i|\psi_{f,k}\rangle_j|^2 \langle\psi_{f,k}|_j \hat{A}_j|\Psi_i\rangle/\langle\psi_{f,k}|_j\Psi_i\rangle
\end{equation}
where the complete set of single-member final states $\{|\psi_{f,k}\rangle_{j}\}$ has been inserted. The first term in the summand is the probability that the $j$th member of the ensemble will be found, on performing a strong measurement on $j$, to be in an eigenstate $k$ of the (individual member) observable $\hat{A}$. This probability will, in the limit $N\to\infty$, be the \emph{fraction} of the ensemble that satisfies the generic pre-selection criterion plus the restricted post-selection of the particular eigenstate $k$. It thus defines a pre- and post-selected sub-ensemble. Equation\eqref{Eq_4_} therefore shows that $\langle\hat{A}_j\rangle$ of a single member $j$ may be expressed as a sum of all sub-ensembles (i.e. a sum over all post-selections $k$) of the total ensemble, with the relative fractions of the ensemble weighted by the corresponding quantities

\begin{equation}\label{Eq_5_}
\langle \psi_{f,k}|_j \hat{A}_j|\Psi_i\rangle/\langle\psi_{f,k}|_j\Psi_i\rangle = \langle\psi_{f,k}|_j \hat{A}_j|\psi_i \rangle_j / \langle \psi_{f,k}|_j \psi_i \rangle_j
\end{equation}
Once again, because all members of the ensemble are identical, the index $j$ is superfluous and will be dropped. The quantity 

\begin{equation}\label{Eq_6n_}
\langle \psi_{f,k}|\hat{A}|\psi_i\rangle/\langle \psi_{f,k}|\psi_i\rangle
\end{equation}
is known as the \emph{weak value} of the operator, evaluated in this case for the specific post-selected eigenstate $k$. More generally, we may define the weak value of an operator $\hat{A}$ as

\begin{equation}\label{Eq_7n_}
w = \langle final | \hat{A} | initial \rangle / \langle final | initial \rangle
\end{equation}
for generic pre-selected initial and post-selected final states $| initial \rangle$ and $| final \rangle$ respectively. Weak values for pre- and post-selected ensembles are not expectation values; rather, they are components of expectation values. They may lie outside the range of eigenvalues, and may even be complex, although the probability of post-selection yielding highly unusual weak values is normally very small \cite{bib4}.  Weak values have been the subject of a considerable theoretical and experimental literature. Good summaries of the interpretational aspects have been provided by Aharonov and Vaidman \cite{bib3b}, and Aharonov and Rohrlich \cite{bib5}, and will not be repeated here. The subject of weak values has stimulated research in topics as diverse as quantum tunneling \cite{bib6} and super-oscillations \cite{bib7}. It has also led to new experimental tests of quantum mechanics \cite{bib8a,bib8b}.  In this paper I shall focus on a specific and important example of weak measurements and post-selection, namely, the decay of an unstable system.

Consider an ensemble of quantum systems prepared at time $t_i$ in identical quantum states $|\psi_i\rangle$, and a sub-ensemble that is post-selected in state $|\psi_f\rangle$ at time $t_f$. The quantum weak value of an observable with Hermitian operator $\hat{A}$ is, at time $t$, $t_i\leq t \leq t_f$, given by

\begin{equation} \label{Eq_1o_} 
w = \frac{\langle \psi_f | U^\dagger (t-t_f) \hat{A} U (t-t_i) |\psi_i\rangle}{\langle\psi_f| U^\dagger (t-t_f) U (t- t_i) |\psi_i\rangle} ,
\end{equation} 
where $U$ is the evolution operator for the system

\begin{equation} \label{Eq_2o_} 
U(t)= e^{-iHt} 
\end{equation} 
and \textit{H} is the Hamiltonian. Most applications of Eq.\eqref{Eq_1o_} do not involve time-dependent evolution in the interval $t_i\leq t\leq t_f$. In this paper I shall consider explicit time dependence. (A rather different approach to time-dependent weak values involving weak energies of evolution has been considered by Parks \cite{bib9}). 

A very basic property of quantum systems is the exponential decay law, where the probability that a given system in an excited state at $t_i$ remains in an excited state after a time $t$ is 

\begin{equation} \label{Eq_3o_} 
P(t)=e^{-2\gamma (t-t_i)} ,
\end{equation} 
where the decay constant, 2$\gamma $, is determined by the interaction strength between the system of interest and a set of ``receptor'' systems to which it is coupled. For example, the system might be an excited atom and the receptor system a bath of simple harmonic oscillators representing modes of the electromagnetic field. 

Equation \eqref{Eq_3o_} is interpreted as follows: if a measurement is made on an individual system at time $t$, then the probability of finding it in the excited state is given by $P(t)$. The measurement is understood to be a strong, projective, measurement, following which the state of the system is ``reset'' to $|\psi_i\rangle$ if it is indeed found that decay has not occurred. The question now arises of what will be found if a weak measurement is performed at time $t$.  In the absence of post-selection, one would expect to recover the same exponential decay law. However, suppose that a post-selection were made at time $t_f$; for example, consider the sub-ensemble of systems that have definitely decayed at time $t_f$. What would be the decay law replacing Eq.\eqref{Eq_3o_}, satisfying the dual constraints

\begin{equation} \label{Eq_4o_} 
P(t_i)= 1 
\end{equation} 

\begin{equation} \label{Eq_5o_} 
P(t_f)= 0\ ?
\end{equation} 

To derive this weak measurement post-selection decay law, I shall consider an excited two-level atom 0 coupled to a large bath of other two-level atoms, initially in their ground states, and compute the weak expectation value of the projection operator onto the excited state of 0 at time $t$, subject to constraints \eqref{Eq_4o_} and \eqref{Eq_5o_}.

It is helpful to illustrate the calculation by first considering a simpler time-dependent system.


\section{Time-dependent weak values}

Consider an electron of charge $e$ at rest in a magnetic field $\mathbf{B}$. The interaction Hamiltonian is

\begin{equation} \label{Eq_6_}
H= -\boldsymbol{\mu} \centerdot \mathbf{B},
\end{equation}
where

\begin{equation} \label{Eq_7_}
\boldsymbol{\mu}=\ -e\hbar\mathbf{S}/m
\end{equation} 

\begin{equation} \label{Eq_8_}
\mathbf{S}=\frac{1}{2} (\sigma_x,\sigma_y,\sigma_z)
\end{equation} 
and $\sigma_i$  are the Pauli spin matrices. Suppose for simplicity that $\mathbf{B}$ lies in the $z$ direction. Then Eq.\eqref{Eq_6_} reduces to 

\begin{equation} \label{Eq_9_} 
H=\hbar \omega \sigma_z,
\end{equation} 
where

\begin{equation} \label{Eq_10_} 
\omega =eB_z/m.
\end{equation} 

The evolution operator \eqref{Eq_2o_} is easily computed by expanding the exponential, using the relation

\begin{equation} \label{Eq_11_} 
\sigma^2_z=1,
\end{equation} 
and summing, to find

\begin{equation} \label{Eq_12_} 
U(t)= \left[ \begin{matrix}
e^{i\omega t/2} & 0 \\ 
0 &  e^{-i\omega t/2} \end{matrix}
\right] 
\end{equation} 
from which one immediately verifies unitarity 

\begin{equation} \label{Eq_13_} 
UU^{\dagger }=U^{\dagger }U=\mathbb{I} 
\end{equation} 
and the evolution property

\begin{equation} \label{Eq_14_} 
U(t_1-t_2)U(t_2-t_3)= U(t_1-t_3).
\end{equation} 

Suppose at initial time $t_i$ the spin points in the +$x$ direction, i.e. $|\psi_i\rangle$ is the normalized eigenstate of the spin operator $\sigma_x$ with eigenvalue +1:

\begin{equation} \label{Eq_15_}
|\psi_i\rangle = \frac{1}{\sqrt{2}}\begin{pmatrix} 1 \\ 1 \end{pmatrix}
\end{equation}

\begin{equation} \label{Eq_16_}
 \sigma_x |\psi_i\rangle = +|\psi_i\rangle.
\end{equation}

The projection operator onto the eigenstate \eqref{Eq_15_} is

\begin{equation} \label{Eq_17_} 
P_{x+}=\frac{1}{\sqrt{2}} \begin{pmatrix} 1&1\\1&1 \end{pmatrix}.
\end{equation} 

The expectation value of $P_{x+}$ at time $t \geq t_i$ is then

\begin{equation} \label{Eq_18_}
\begin{split}
\langle\psi_i| P_{x+}(t)|\psi_i\rangle &\equiv \langle\psi_i| U^\dagger (t-t_i) P_{x+} U(t-t_i) | \psi_i \rangle \\  &= \frac{1}{2} [ 1+\cos \omega (t-t_i)].
\end{split}
\end{equation}

The projection operator onto the +1 eigenstate of $\sigma_y$ is

\begin{equation} \label{Eq_19_} 
P_{y+}=\frac{1}{2}\begin{pmatrix} 1&-i\\i&1 \end{pmatrix},
\end{equation} 
from which it follows that 

\begin{equation} \label{Eq_20_}
\begin{split}
 \langle \psi_i |P_{y+}(t)| \psi_i \rangle &\equiv \langle \psi_i | U^\dagger(t-t_i) P_{y+} U(t-t_i)| \psi_i\rangle \\ &= \frac{1}{2} [1-\sin \omega(t-t_i)].
\end{split}
\end{equation}

From Eqs.\eqref{Eq_18_} and \eqref{Eq_20_} one may deduce the well-known result that the spin direction precesses with frequency $\omega $ in the \textit{x-y} plane.

The foregoing results may now be compared with weak measurements made at time $t$, with initial state $|\psi_i \rangle$ given by Eq.\eqref{Eq_15_}, and various choices of post-selection.  For example, with

\begin{equation} \label{Eq_21_} 
|\psi_f\rangle =\frac{1}{\sqrt{2}}\begin{pmatrix} i\\-1 \end{pmatrix}, \ \frac{1}{\sqrt{2}}\begin{pmatrix} 1\\-1 \end{pmatrix}, \frac{1}{\sqrt{2}}\begin{pmatrix} 1\\1 \end{pmatrix}
\end{equation} 
corresponding to the eigenstates along the $+y$, $-x$ and the $+x$ axes respectively, it follows using Eqs.\eqref{Eq_1o_} and \eqref{Eq_12_} that at time $t$, $t_i\leq t\leq t_f$, the weak values $w(P_{x+})$ of the projection operator $P_{x+}$ are 

\begin{equation}\label{Eq_22_}
 \cos \dfrac{\omega(t-t_i)}{2} \dfrac{\left[\cos \dfrac{\omega(t_f-t)}{2} + \sin \dfrac{\omega(t_f-t)}{2} \right]}{\left[ \cos \dfrac{\omega(t_f-t_i)}{2} - \sin \dfrac{\omega(t_f-t_i)}{2} \right]}
\end{equation}

\begin{equation} \label{Eq_23_} 
\frac{1}{2}+\ \frac{\sin [\omega(2t-t_i-t_f)/2]}{2\sin[\omega(t_f-t_i)/2]}
\end{equation} 

\begin{equation} \label{Eq_24_} 
\frac{\cos[\omega(t-t_i)/2] \cos[\omega(t_f-t)/2]}{\cos \omega(t_f-t_i)/2} 
\end{equation} 
respectively. The corresponding weak values for $P_{x-}$  may be derived in from the fact that, for the same choices of $|\psi_i\rangle$ and $|\psi_f\rangle$,

\begin{equation} \label{Eq_25_}
 w(P_{x+})+w(P_{x-})=1.
\end{equation}
It is also straightforward to calculate weak values for $P_{y\pm}$, $\sigma_x$ and $\sigma_y$.

 An important check on \eqref{Eq_24_} is to set

\begin{equation} \label{Eq_26_}
 \omega(t_f-t_i) = 2n \pi, \ n=1,2,3,\ldots
\end{equation}
corresponding to an integral number of precession cycles, for then $|\psi_f\rangle=|\psi_i\rangle$, and post-selection is trivial. Thus, applying condition \eqref{Eq_26_} reduces \eqref{Eq_24_} to Eq.\eqref{Eq_18_} as required. Similarly, if 

\begin{equation} \label{Eq_27_}
 \omega (t_f-t_i) = \dfrac{\pi}{2}
\end{equation}
then $|\psi_f\rangle$ corresponds to the $-x$ eigenvalue, and \eqref{Eq_23_} reduces to 

\begin{equation} \label{Eq_28_}
 \dfrac{1}{2} [ 1- \cos \omega(t-t_i)]
\end{equation}
which is easily shown to correspond to the strong expectation value

\begin{equation} \label{Eq_29_} 
\langle \psi_i | P_{x-} (t) | \psi_i \rangle \equiv \langle \psi_i | U^\dagger (t-t_i) P_{x-} U(t-t_i) | \psi_i \rangle.
\end{equation} 

Thus the weak and strong expectation values coincide for all $t$ in the interval $t_i \leq t \leq t_f$ when $|\psi_f\rangle$ is chosen to be the eigenstate that the system would reach at time $t_f$ under unitary evolution from the initial state $|\psi_i\rangle$. The result follows from the general property that

\begin{equation} \label{Eq_30_} 
\begin{split}
&\frac{\langle\psi_f|U^\dagger(t-t_f)\hat{A}U(t-t_i)|\psi_i\rangle}{\langle\psi_f|U(t_f - t_i)|\psi_i\rangle}\\ &\qquad \qquad \qquad =  \langle \psi_i | U^\dagger (t-t_i) \hat{A}U (t-t_i)| \psi_i \rangle
\end{split}
\end{equation} 
when 

\begin{equation} \label{Eq_31_}
 |\psi_f\rangle=U(t_f-t_i)|\psi_i\rangle
\end{equation}
which follows immediately by using Eqs.\eqref{Eq_13_} and \eqref{Eq_14_}.

 Expressions \eqref{Eq_22_}-\eqref{Eq_24_} display some curious properties typical of weak values. The weak expectation value for $P$ can lie outside the normal range of probabilities [0,1]. It can, for example, take on both negative values, or values $>1$, for appropriate choices of $t_f$, $t_i$ and $t$. Special interest attaches to \eqref{Eq_24_}, which is the weak expectation value of the projection operator onto the eigenstate of spin in the $+x$ direction, when this state is chosen for both pre-selection and post-selection. Naively, one might have supposed that $w(P_{x+})$ would be fixed at 1 throughout, but in fact it is time-dependent. By way of illustration, if we choose condition \eqref{Eq_27_}, corresponding to a quarter of a precession period, then 

\begin{equation} \label{Eq_32_}
 w(P_{x+}) = \frac{1}{2} [ 1+ \sin \omega(t-t_i) + \cos \omega (t-t_i)],
\end{equation}
which takes the required value of 1 at $t=t_i$ and $t=t_f$, but in between is time-dependent and $>1$.


\section{Decay Law}

 I now come to the case of the decay of an excited state by considering an initially excited two-level atom coupled to a bath of $2N$ other two-level atoms initially in their ground states. The atoms are labeled by the index $n$. It is convenient to choose the ground state energies of all the atoms to coincide and be set to zero, and the excited states $E_n$ to satisfy the relation

\begin{equation} \label{Eq_33_}
 E_n - E_0 = n \Delta E, \ -N \leq n \leq N,
\end{equation}
i.e. the excited states are equispaced and distributed symmetrically about the excited state of the reference atom, labeled by $n=0$. For simplicity I assume that the reference atom is coupled equally to each atom of the bath, and the interaction is described by the real constant Hamiltonian $H$.

 The Schr\"{o}dinger equation for this system is equivalent to the coupled differential equations

\begin{equation} \label{Eq_34_} 
\dot{a}_0= -i\sum_n Ha_ne^{-in\Delta Et}
\end{equation} 

\begin{equation} \label{Eq_35_} 
\dot{a}_n= -iHa_0e^{in\Delta Et} 
\end{equation} 
where $a_n$ is the amplitude of the excited state of atom $n$, and here and henceforth I set $\hbar = 1$. Equations \eqref{Eq_34_} and \eqref{Eq_35_} may be solved exactly (i.e. without the use of perturbation theory) with the method of Laplace transforms \cite{bib10}. One finds

\begin{equation} \label{Eq_36_} 
a_0(t)= \mathcal{L}^{-1} \left[\dfrac{1}{s+\frac{\pi H^2}{\Delta E}\coth\left(\frac{\pi s}{\Delta E}\right)}\right].
\end{equation} 
The inverse transform $\mathcal{L}^{-1}$ may be evaluated explicitly for $n=0$ in the limiting case $N\to \infty$,  $\Delta E\to 0$, $H\to 0$ such that  

\begin{equation} \label{Eq_37_} 
\frac{H^2\pi}{\Delta E}\to \gamma
\end{equation} 
where $\gamma$ is a finite constant and use has been made of the summation relation \eqref{Eq_A.1_} in Appendix A. Then

\begin{equation} \label{Eq_38_} 
a_0(t)= e^{-\gamma (t-t_i)}.
\end{equation} 
Note that Eq.\eqref{Eq_38_} is an exact solution (in the limit \eqref{Eq_37_}) for the amplitude of $a_0(t)$. It follows that 

\begin{equation} \label{Eq_39_} 
|a_0(t)|^2=e^{-2\gamma(t-t_i)}
\end{equation} 
coinciding with the usual exponential decay law Eq.\eqref{Eq_3_} for the probability of finding atom 0 in the excited state on performing a strong, projective measurement at time $t$. It has been derived here without the use of phase randomization or perturbation approximation. The foregoing analysis is thus ideally suited to generalization to weak values.

To apply Eq.\eqref{Eq_1o_} we need to calculate the evolution operator $U(t)$ for the system described by Eqs.\eqref{Eq_34_} and \eqref{Eq_35_}. Restricting to the situation where only one atom at a time is excited, the evolution operator $U(t)$ for the relevant subspace of the full Hilbert space of states will be a $(2N+1)\times (2N+1)$ matrix, the components of which may be calculated from the solutions of \eqref{Eq_34_} and \eqref{Eq_35_}. From Eq.\eqref{Eq_38_} we have 

\begin{equation} \label{Eq_40_} 
U_{00}(t)= e^{-\gamma t} 
\end{equation} 
in the limit $\Delta E\to 0$. Using this limiting solution, Eqs.\eqref{Eq_34_} and \eqref{Eq_35_} may be solved to give

\begin{equation} \label{Eq_41_} 
U_{n0}(t)=i H\left[\dfrac{e^{-\gamma t+in\Delta E t}-1}{\gamma - i n \Delta E}\right],
\end{equation} 
which is also understood as to be taken in the limit $\Delta E \to 0$. The other components of $U_{nm}(t)$ may be solved similarly, after some labor, but are not needed for the purposes of this calculation.

 Equation \eqref{Eq_1o_} may be rewritten, using relation \eqref{Eq_13_} plus

\begin{equation} \label{Eq_42_} 
U^\dagger(t)= U(-t) 
\end{equation} 
as follows

\begin{equation} \label{Eq_43_} 
w= \frac{\langle\psi_f|U(t_f-t)\hat{A}U(t-t_i)|\psi_i\rangle}{\langle\psi_f|U(t_f-t_i)|\psi\rangle}.
\end{equation} 

If the operator $\hat{A}$ is chosen to be the projection operator $P_{\uparrow}$ onto the excited state of atom 0, then Eq.\eqref{Eq_43_} describes the weak value for finding the atom undecayed at time $t$, given the condition that it is prepared in the excited state at time $t_i$ and post-selected to have decayed at time $t_f$. A possible choice of final state is 

\begin{equation} \label{Eq_44_} 
|\psi_f\rangle =|\psi_k\rangle 
\end{equation} 
where the $k$th atom only in the bath is excited, and the reference atom 0 is in the ground state. In a physically realistic scenario, this corresponds to the atom in the ground state and a photon of energy $E_k=k\Delta E$ having been emitted.

To evaluate \eqref{Eq_43_} for this choice, we note that $P_{\uparrow }$ is represented by a $(2N+1)\times (2N+1)$ matrix with all components 0 except

\begin{equation} \label{Eq_45_} 
(P_{\uparrow})_{00}=1.
\end{equation} 

The state $\langle\psi_k|$ is represented by a row vector with all components 0 except the $k$th, which is 1 (for normalization). Similarly $|\psi_i\rangle$ is a column vector with the $i=0$ component 1 and all others 0. Then equation \eqref{Eq_43_} reduces to

\begin{equation} \label{Eq_46_} 
w=\frac{U_{k0}(t_f-t)U_{00}(t-t_i)}{U_{k0}(t_f-t_i)}.
\end{equation} 

Using \eqref{Eq_40_} and \eqref{Eq_41_} we find 

\begin{equation} \label{Eq_47_} 
w= e^{-\gamma (t-t_i)}\left[\dfrac{1-e^{-\gamma (t_f-t)+ i (E_k-E_0)(t_f-t)}}{1-e^{-\gamma (t_f-t_i)+i (E_n-E_0) (t_f-t_i)}}\right].
\end{equation} 
The fact that $w$ is complex is a familiar feature of weak values \cite{bib5}; both the real and imaginary parts have physical interpretations. In the case that $E_k = E_0$, i.e. the ``photon'' energy coincides with the reference atom's excitation energy, then \eqref{Eq_47_} reduces to the simple expression

\begin{equation} \label{Eq_48_} 
w= e^{-\gamma (t-t_i)}\left[\dfrac{1-e^{- \gamma(t_f-t)}}{1-e^{-\gamma (t_f-t_i)}}\right].
\end{equation} 


Equation \eqref{Eq_48_} generalizes the familiar exponential decay law \eqref{Eq_38_} to the case of weak measurements.  Note that

\begin{equation} \label{Eq_49_} 
 \begin{array}{ccccccc}
w & = & 1 & \qquad \ \qquad & t & = & t_i \\
& = & 0 & \qquad \ \qquad & t & = & t_f
  \end{array}
\end{equation}
as required, given that the reference atom is known to be in the excited state at $t=t_i$ and to be in the ground state at $t=t_f$. Furthermore, as $t_f\to \infty $, \eqref{Eq_48_} reduces to the standard exponential decay law \eqref{Eq_38_}, which is easy to understand because the system is known to have decayed in the limit $t\to \infty $, so post-selection of the ground state is redundant.

Equation \eqref{Eq_43_} may be evaluated for other choices of operator $\hat{A}$ and final states $|\psi_f\rangle$. One case of interest is $\hat{A}=P_{\uparrow }$ and the $k$th component of $\langle\psi_k|$ given by

\begin{equation} \label{Eq_50_} 
\langle\psi_f|_k=\frac{iH}{\gamma +ik\Delta E},\ \forall k.
\end{equation} 
Inspection of Eq.\eqref{Eq_41_} shows that this is the state to which the system approaches asymptotically as $t\to \infty$, being a superposition of all excitations of bath atoms. It corresponds to the emission by atom $0$ of all possible photon energies. If this state is post-selected at the finite time $t_f$, one must replace \eqref{Eq_46_} by

\begin{equation} \label{Eq_51_} 
\begin{split}
w &= U_{00}(t-t_i)\sum^{\infty }_{k=-\infty} \frac{U_{k0}(t_f-t)}{\gamma +ik\Delta E} / \sum^{\infty }_{k=-\infty } \frac{U_{k0}(t_f-t_i)}{\gamma +ik\Delta E} \\ & {\underset{\Delta E\to 0}{\longrightarrow}} e^{-\gamma (t-t_i)}\left[\frac{1-e^{-2\gamma (t_f-t)}}{1-e^{-2\gamma (t_f-t_i)}}\right] 
\end{split}
\end{equation} 
which also satisfies \eqref{Eq_49_} and the condition that it reduces to \eqref{Eq_38_} when $t_f\to \infty$. To arrive at Eq.\eqref{Eq_51_} I have made use of Eqs.\eqref{Eq_37_}, and \eqref{Eq_A.1_} and \eqref{Eq_A.2_} in Appendix A. 

Finally, I note that if one chooses $|\psi_f\rangle = |\psi_i\rangle$, then 

\begin{equation} \label{Eq_52_} 
w(P_{\uparrow})=\frac{U_{00}(t_f-t)U_{00}(t-t_i)}{U(t_f-t_i)}=1 
\end{equation} 
in contrast to the time-dependent result \eqref{Eq_24_} in the spin-precession case.  In spite of this result, Eq.\eqref{Eq_52_} does not imply that weak measurements of the bath atoms will show them remaining inert in this case.  If we denote the corresponding projection operators for the bath atoms by $P_{\uparrow n}$, then 

\begin{equation}\label{Eq_53_}
w(P_{\uparrow n}) \neq 0.
\end{equation}
However, unitarity plus Eq.\eqref{Eq_52_} implies

\begin{equation}\label{Eq_54_}
\sum_n w(P_{\uparrow n}) = 0
\end{equation}
where the summation is taken over all bath atoms.  Equations \eqref{Eq_53_} and \eqref{Eq_54_} are consistent only if the summand contains both positive and negative contributions.  The appearance of negative weak values for a projection operator is another unusual but familiar aspect of weak measurement theory, as remarked in section 2.  A full discussion of this topic will be presented in a future paper.


\section{Discussion}

The exponential decay law is one of the most fundamental results of quantum mechanics, with wide applicability to atomic and nuclear physics. Concealed in the textbook discussion, however, is the fact that the exponential law \eqref{Eq_3_} is actually a restricted case of a more general result for quantum de-excitation. The standard exponential result assumes that the system is inspected via a strong, projective measurement. By considering weak measurements the exponential law can be generalized, for example to \eqref{Eq_48_} and \eqref{Eq_51_}, depending on the choice of post-selection.

Post-selection of sub-ensembles of atoms that are in their ground states at time $t$ is conceptually straightforward, and amenable to experimental test.


\appendix
\section{}

The right-hand side of Eq.\eqref{Eq_51_} involves the summations

\begin{equation}\label{Eq_A.1_}
 \lim_{\Delta E \to 0} \Delta E \sum_{k=-\infty}^\infty \dfrac{1}{\gamma^2+k^2\Delta E^2} = \dfrac{\pi}{\gamma}
\end{equation}

\begin{equation} \label{Eq_A.2_}
 \lim_{\Delta E \to 0} \Delta E \sum_{k=-\infty}^\infty \dfrac{e^{ik\Delta Et}}{\gamma^2 + k^2\Delta E^2} = \dfrac{\pi}{\gamma}e^{-\gamma t} \ (t>0)
\end{equation}

Equation\eqref{Eq_A.1_} follows immediately from the relation

\begin{equation}\label{Eq_A.3_}
 \Delta E \sum_{k=-\infty}^\infty \dfrac{1}{\gamma^2 + k^2 \gamma \Delta E^2} = \dfrac{\pi}{\gamma} \coth \left(\dfrac{\pi \gamma}{\Delta E}\right)
\end{equation}
(see, for example \cite{bib11}). Equation\eqref{Eq_A.2_} may be proved as follows:

\begin{equation}\label{Eq_A.4_}
 \begin{split}
  \Delta E \sum_{k=-\infty}^\infty & \dfrac{e^{ik\Delta Et}}{\gamma^2 + k^2 \Delta E^2} = \dfrac{2}{\Delta E} \sum_{k=1}^\infty \dfrac{\cosh \Delta Et}{k^2+\gamma^2/\Delta E^2}-\dfrac{\Delta E}{\gamma^2} \\ &= \dfrac{\pi}{\gamma} \dfrac{\cosh[(\pi-\Delta Et)\gamma/\Delta E]}{\sinh (\pi \gamma/\Delta E)} - \dfrac{\Delta E}{\gamma^2} \\ &= \dfrac{\pi}{\gamma} [ \coth (\dfrac{\pi\gamma}{\Delta E} \cosh \gamma t - \sinh \gamma t ] -\dfrac{\Delta E}{\gamma^2}
 \end{split}
\end{equation}

The summation on the first line of Eq.\eqref{Eq_A.4_} may be found, for example, in \cite{bib12}. Finally, taking the limit $\Delta E\to 0$, Eq.\eqref{Eq_A.2_} follows.


\acknowledgements

I should like to thank Yakir Aharonov for assistance and encouragement, Robert Griffiths and Allen D. Parks for helpful comments, and Jeffrey Tollaksen for extensive assistance with the background theory.

\bibliography{paper.bib}

\end{document}